\documentclass[usegraphicx,useAMS]{mn2e}
\usepackage{times}
\usepackage{amsmath}

\newcommand{\cms}{\,{\rm cm$^{-2}$}\,}

\newcommand{\kmsmpc}{\,{\rm km\,s$^{-1}$\,Mpc$^{-1}$}\,}
\newcommand{\cmc}{\,{\rm cm$^{-3}$}\,}

\newcommand{\kel}{\,{\rm K\ }}
\newcommand{\etal}{{ et~al.~}}

\newcommand{\LCDM}{$\Lambda$CDM\ }
\newcommand{\HH}{H$_2$\ }
\newcommand{\HI}{\hbox{H~$\rm \scriptstyle I\ $}}

\newcommand{\HeI}{\hbox{He~$\rm \scriptstyle I\ $}}
\newcommand{\HeII}{\hbox{He~$\rm \scriptstyle II\ $}}

\newcommand{\fluxunit}{\,{\rm erg\,s$^{-1}$\,cm$^{-2}$\,Hz$^{-1}$}\,}
\newcommand{\Lya}{Ly$\alpha\ $}
\newcommand{\Ms}{M_\odot}
\newcommand{\MsMpc}{M_\odot\,\hbox{\rm Mpc$^{-3}$}}

\begin{document}


\title{Effects of a Soft X-ray Background on Structure Formation
   at High Redshift}

\author[Machacek, Bryan \& Abel]{
M.E.~Machacek,$^1$\thanks{mariem@space.mit.edu, on leave from
Department of Physics, Northeastern University, Boston, MA 02115}
G.L.~Bryan,$^2$\thanks{gbryan@astro.ox.ac.uk}
T.~Abel,$^3$\thanks{hi@tomabel.com} \\
$^1$
Center for Space Research, Massachusetts Institute of Technology,
       70 Vassar St., Cambridge, MA 02139, USA \\
$^2$
Department of Physics, Oxford University, Keble Road, Oxford,
   OX1 3RH, UK \\
$^3$
Department of Astronomy and Astrophysics,The Pennsylvania State University, 525 Davey Lab, University Park, PA 16802, USA}

\maketitle


\begin{abstract}

We use three dimensional hydrodynamic simulations to investigate the effects
of a soft X-ray background, that could have been
produced by an early generation of mini-quasars, on the subsequent cooling and
collapse of high redshift pregalactic clouds.  The simulations use an
Eulerian adaptive mesh refinement technique with initial conditions drawn
from a flat $\Lambda$-dominated cold dark matter model cosmology to follow the
nonequilibrium chemistry of nine chemical species in the presence of both a
soft ultraviolet Lyman-Werner \HH photodissociating flux of strength
$F_{LW}=10^{-21}$~\fluxunit and soft X-ray background extending to $7.2$~keV
including the ionization and heating effects due to secondary electrons.
Although we vary the normalization of the X-ray background by two orders of
magnitude, the positive feedback effect of the X-rays on cooling and collapse
of the pregalactic cloud expected due to the increased electron fraction is
quite mild, only weakly affecting the mass
threshold for collapse and the fraction of gas within the cloud that is able
to cool, condense and become available for star formation.
Inside most of the
cloud we find that \HH is in photodissociation equilibrium with
the soft UV flux. The net buildup of the electron density needed to
enhance \HH formation occurs too slowly compared to the \HH photodissociation
and dynamical timescales within the cloud to overcome the
negative impact of the soft UV photodissociating flux on cloud collapse.
However, we find that even in the most extreme cases the first objects to 
form do rely on molecular hydrogen as coolant and stress that our results do 
not justify the neglect of these objects in models of galaxy formation.
Outside the cloud we find the dominant effect of a sufficiently
strong X-ray background is to heat and partially ionize the inter-galactic
medium, in qualitative agreement with previous studies.
\end{abstract}

\begin{keywords}
cosmology:theory -- early universe -- galaxies:formation
\end{keywords}


\section{Introduction}
\label{sec:introduction}

One of the most important questions in current cosmology is to understand
how the cosmological dark ages ended by identifying the nature of
the first luminous sources and determining their impact on subsequent
structure formation and the reionization of the universe.
Recent observations are
beginning to constrain the epoch of reionization and give modest information
about possible first sources. Observations of metals throughout even low
column density \Lya lines (Ellison \etal 1999, 2000; Schaye \etal 2000)
suggest the need for an early population of stars to pre-enrich the
inter-galactic medium (IGM).
Observed spectra of high redshift quasars such as SDSS 1030+0524
(Fan \etal 2001) at $z=6.28$ and \Lya emitters at these redshifts
(e.g. Hu \etal 2002) start to constrain the overlap stage of
hydrogen reionization. Observations of patchiness in the HeII
optical depth in the Lyman alpha forest at $z \sim 3$ (Reimers
\etal 1997) might signal a recent period of helium reionization.
Multi-wavelength observations planned in the near future hope to
study the reionization epoch in detail. Direct imaging of quasars
and star clusters at $z > 10$ may be possible using the Next
Generation Space Telescope (Haiman \& Loeb 1999, Barkana \& Loeb
2000). Emission measurements in the 21 cm line using LOFAR and
the Square Kilometer Array could identify the first epoch of
massive star formation (Tozzi \etal 2000). Searches for
polarization effects and secondary anisotropies in the Cosmic
Microwave Background induced by scattering off the increased
electron fraction produced during reionization are planned by
Planck and the next generation of millimeter telescopes such as
ALMA.

In order to correctly interpret the findings of these observations,
we need to understand the predictions of the current cosmological
structure formation paradigm.
The initial stages of collapse of structure at
high redshift in cold dark matter cosmologies have been well studied
analytically and numerically (see, for example, the excellent review by
Barkana \& Loeb 2001 and references therein). Structure forms from small
density perturbations via
gravitational instability where smaller clumps merge to form larger clumps
within and at the intersections of filaments. By redshifts
$30 \la z \la 20$ pregalactic clouds with masses
(virial temperatures) $M  \sim 10^5$ - $10^6\,\Ms$
( $T_{vir} \sim 1000$~\kel) have formed sufficient molecular
hydrogen to begin to cool reducing pressure support in their
central regions and allowing their cores to collapse to high
density
(Haiman, Thoul \& Loeb 1996; Tegmark \etal 1997; Omukai \& Nishi 1998; 
 Abel \etal 1998; Fuller \& Couchman 2000). 
However, it is only recently that 3-D numerical
simulations have achieved sufficient resolution to follow this
collapse reliably from cosmological initial conditions to the
extremely high densities and stellar spatial scales expected for
the first luminous sources (Abel, Bryan, \& Norman 2000, 2001).
These  and related studies (e.g. Bromm, Coppi \& Larson 2001;
Nakamura \& Umemura 2001) show that the final stages of collapse
are slow (quasi-static) and that fragmentation in the primordial,
metal-free gas is difficult. Thus the first sources were most
probably 
massive stars. This has led to renewed interest in
the evolutionary properties of such metal-free, 
massive stars
(Fryer, Woosley \& Heger 2001; Schneider \etal 2001; Oh 2001; Oh
\etal 2001).
However, the micro-galaxies ($\ga 10^6\Ms \approx 10^{-6}$ times the
mass of the Milky Way) which  host the first stars collapse over a
large redshift interval. At the same time in different regions of
space also rarer but much larger objects are forming whose integrated
emitted light cannot be reliably predicted nor constrained strongly from
existing observations. Consequently, the radiation spectrum from the
first luminous sources is uncertain. 

Clearly, a background of soft ultraviolet (UV) radiation is expected
from the first stars. The neutral primordial gas remains optically
thin to soft UV photons below
the hydrogen ionization edge until the gas collapses to high density. In
particular photons in the Lyman-Werner bands
($11.2\,{\rm eV}\, < E_\gamma < 13.6$~eV)
can travel large distances and readily photodissociate the fragile \HH
coolant in their own and neighboring clouds via the two-step Solomon process.
Such a soft UV background alone is expected to suppress the
subsequent collapse of low mass clouds
($T_{vir} < 10^4$~\kel) that require \HH to cool (Dekel \& Rees 1987;
Haiman, Rees \& Loeb 1997; Omukai \& Nishi 1999; Ciardi \etal 2000;
Haiman, Abel \& Rees 2000; Glover \& Brand 2001;
Machacek, Bryan \& Abel 2001; Oh \& Haiman 2001).
In Machacek, Bryan \& Abel (2001, hereafter referred to as MBAI) we used
fully 3-dimensional AMR simulations 
in the optically thin approximation 
starting from cosmological initial
conditions in a \LCDM cosmology to follow the evolution and collapse
of pregalactic clouds in the presence of varying levels of soft UV flux,
$0 \leq F_{LW} \leq 10^{-21}$~\fluxunit. We confirmed that the presence
of a  Lyman-Werner flux $F_{LW}$ delays the onset of collapse  until the
pregalactic clouds evolve to larger masses and found a fitting
formula for the mass threshold for collapse given the mean flux in the
Lyman-Werner bands. In MBAI we also investigated what fraction of gas could
cool and condense and thus become available for star formation, an important
input parameter for semi-analytical models of galaxy formation, stellar
feedback and its impact on the process of reionization
(Ciardi \etal 2000, Ciardi, Ferrara, \& Abel 2000, 
Madau, Ferrara, \& Rees 2001).
We found that the fraction of gas that could cool and become
dense in metal-free pregalactic clouds in our simulations depended primarily on
two numbers, the flux of soft UV radiation and the mass of the cloud, and
that once above the collapse mass threshold determined by the level of
photodissociating flux, the fraction of cold, dense gas available for star
formation in these early structures increased logarithmically with the cloud's
mass.

If early luminous sources include a population of mini-quasars or
other X-ray emitting sources such as X-ray binaries they will produce
a background radiation field of soft X-rays with energies above the
Lyman limit ($\ga 1$~keV).  It has been suggested (Haiman, Rees \&
Loeb 1996; Haiman, Abel \& Rees 2000; Oh 2001; Ricotti, Gnedin \&
Shull 2001) that the increased electron fraction produced by the
ionizing photons would promote the formation of \HH thereby undoing
the negative feedback effect of the soft, Lyman-Werner UV flux on the
collapse of low mass pregalactic clouds.  Haiman, Rees
\& Loeb (1996) found that the formation rate of \HH could be enhanced
in a dense ($n_H \ga 1$~cm$^{-3}$), stationary, homogeneous gas
cloud of primordial composition irradiated by an external, uniform
power-law background flux with photon energies $\la 40$~keV.
Their calculations assumed chemical equilibrium for the species. They
used Lepp \& Shull (1983) cooling functions for molecular hydrogen and
mimicked radiative transfer effects by assuming a mean absorbing
column density of $10^{22}$~\cms. Haiman, Abel \& Rees (2000) again
considered \HH formation in a static, isolated primordial cloud in the
presence of a power-law radiation flux extending to $10$~keV. However,
they adopted the more realistic profile of a truncated isothermal
sphere at its virial temperature and followed the time evolution of
nine chemical species.  They concluded that in the cores of these
objects the negative effect of the photodissociating flux on collapse
is erased if as little as $10\%$ of the radiation field is from
mini-quasars with energies extending into the soft X-ray band.
In two recent papers, Ricotti \etal (2002a, 2002b) have also examined 
the effect of radiative feedback on cooling in low-mass halos and find 
that high-energy photons can have a positive net effect on the star 
formation rate.

In this paper we extend the results of our fully 3-D Eulerian AMR
simulations of the formation and collapse of primordial pregalactic
structure in the presence of a soft Lyman-Werner UV background (MBAI)
to include the contribution of X-rays with energies extending to
$7.2$~keV.  This work, as in MBAI, improves upon earlier studies by
following the time evolution of a collection of collapsing
protogalaxies evolving together in a $1$~Mpc$^3$ (comoving) simulation
volume from cosmological initial conditions drawn from a flat \LCDM
model. Thus it treats consistently the density evolution of the cloud
and includes the effects of gravitational tidal forces and merging
that also impact cooling and collapse. We develop statistics on the
amount of gas that can cool due to molecular hydrogen and the fraction
of gas that is cold and dense enough to be available for star
formation in these objects when exposed to both a soft \HH
photodissociating flux and various levels of X-rays.  We also use the
more recent Galli \& Palla (1998) \HH cooling functions in this work
and fitting functions for the energy deposition from high energy
electrons from Shull \& Van Steenberg (1985) that take into account
the primordial composition of the gas.

This paper is organized in the following way: In  \S\ref{sec:sims} we
review the set-up of our simulations with particular emphasis on our
treatment of photoionization and the effects of secondary electrons
induced by the X-ray background. In \S\ref{sec:data} we discuss our peak
identification method and the general
characteristics of our simulated data set of pregalactic clouds.
In \S\ref{sec:fractions} we investigate how varying the
intensity of the X-ray background affects the amount of gas that can
cool and condense, thus becoming available for star formation, in the
presence of both an \HH photodissociating flux $F_{LW}$ and ionizing
X-ray background.
In \S\ref{sec:profiles} we use radial profiles of cloud properties to
elucidate the effects of the competing physical processes important to
cooling and collapse.
We summarize our results in  \S\ref{sec:conclude}.


\section{Simulations}
\label{sec:sims}

We use the same simulation procedure, cosmology, initial conditions, and
simulation volume as in our previous work (MBAI) so that we can directly
compare the results of both studies. We summarize that technique here for
completeness, but refer the reader to MBAI for a more detailed discussion.
We used a flat, low matter density \LCDM model for the simulations whose
parameters were chosen to give good consistency with observation.
Specifically, $\Omega_0=0.4$, $\Omega_b=0.05$, $\Omega_\Lambda=0.6$, $h=0.65$,
$\sigma_8=0.8$, and $n=1$ where $\Omega_0$, $\Omega_b$,
and $\Omega_\Lambda$ are the fraction of the critical energy density carried
in nonrelativistic matter, baryons, and vacuum energy, respectively, $h$ is
the dimensionless Hubble parameter in units of $100$ \kmsmpc, $\sigma_{8}$
is the density fluctuation normalization in a sphere of radius $8h^{-1}$~Mpc,
and $n$ is the slope of the primordial density perturbation power spectrum.

Our data set consists of results from six simulations starting from
identical cosmological initial conditions on the density fields
derived from the $F_{LW}=10^{-21}$ erg\,s$^{-1}$\,cm$^{-2}$\,Hz$^{-1}$
simulation in MBAI.
That simulation was
initialized at $z=99$ with density perturbations generated for the above
\LCDM model using the Eisenstein \& Hu (1998) transfer functions.
The density perturbations were evolved in a $1$~Mpc$^3$ comoving simulation
volume using a fully 3-dimensional Eulerian adaptive mesh
refinement (AMR) simulation code (Bryan 1999; Bryan \& Norman 1997, 1999)
that forms an adaptive hierarchy of rectangular grid patches at various
levels of resolution where each grid patch covers some region within its
parent grid needing additional refinement and may itself become a parent grid
to an even higher resolution child grid. Once an active region of
structure formation was identified in the simulation box, it was surrounded
by multiple static refinement levels to achieve a mass resolution within that
region of $4.78 \Ms$ and $38.25 \Ms$ in the initial conditions for the gas
and dark matter, respectively. The region of
interest was allowed to dynamically refine further to a total of
$14$ levels on a $64^3$ top grid resulting in a maximum dynamic range of
$10^6$ and comoving spatial resolution at maximum refinement of $0.95$~pc
(corresponding to a physical spatial resolution at $z=19$ of $0.05$~pc).
As in MBAI we call a peak ``collapsed'' when it reaches maximal refinement.
Once a peak has maximally refined we can not follow the evolution of its
innermost region further. At this point we introduced artificial pressure
support within the inner $1 - 2$~pc of the core to stabilize it against
further collapse (and prevent the onset of numerical instability),
so that we could continue to follow the evolution of structure
elsewhere. Since most of the cooling occurs outside of the collapsed
\HH core, this should not affect the determination of the cooled gas
fractions presented in \S\ref{sec:fractions}.

We model the \HH photodissociating Lyman-Werner flux in the same way
as in MBAI by assuming a constant flux $F_{LW}$ throughout the simulation
with mean photon energy $12.86$~eV, thus neglecting its time-dependent
turn-on and buildup.  \HH is photodissociated by the
two-step Solomon process
\begin{equation}
H_2 + \gamma \rightarrow H_2^* \rightarrow H + H ,
\label{eq:solomon}
\end{equation}
with rate coefficient given by (Abel, \etal 1997)
\begin{equation}
k_{diss} = 1.1 \times 10^8 F_{LW}\,\hbox{\rm s$^{-1}$}.
\label{eq:kdiss}
\end{equation}
where $H_2^*$ is any of the $76$ Lyman-Werner resonances in the 
$11.18$--$13.6$~eV energy range.
We work in the optically thin approximation
and argue that the effect of self-shielding of the UV flux
by the collapsing cores will be small so that we can neglect it. We also 
neglect the photodetachment of $H^-$ that might hinder the buildup of \HH since
$H^-$ is needed to catalyze the formation of molecular hydrogen. However, at 
the internal cloud temperatures and densities considered here we expect $H^-$
photodetachment to be suppressed by at least two orders of magnitude relative
to \HH formation. We refer the reader to MBAI, Section 6, for a detailed 
discussion of these neglected processes.

In three of our simulations we turn on an additional ionizing X-ray radiation
field at $z=30$. This redshift was chosen because it was the earliest redshift
found in MBAI at which sources reached maximal refinement in the absence of
any radiation field. For ease of comparison we model the X-ray field
$F_x$ in the same way as Haiman, Abel \& Rees (2000) using an absorbed power
law spectral form given by
\begin{equation}
F_x = \epsilon_x F_{LW} \biggl (\frac {\nu}{12.86} \biggr )^{-\alpha} {\rm exp}[-10^{22}(\sigma_{\HI} + 0.08\sigma_{\HeI})]
\label{eq:spectrum}
\end{equation}
with $\alpha=1$, $F_{LW}=10^{-21}$~\fluxunit, corresponding roughly to
a spectral intensity $J_{21}=0.1$ (Haiman, Abel \& Rees, 2000), and photon
energies extending to $7.2$~keV.
The exponential factor was
chosen to approximate the absorption of photons with energies
above $13.6$~eV by the
neutral IGM 
(and thus mimic the effects of radiative transfer on average) 
by assuming a constant absorbing column density of $10^{22}$~\cms
for hydrogen, a helium column density reduced by a factor $0.08$, representing
the ratio of helium to hydrogen number densities, and ionization cross
sections $\sigma_{HI}$ and $\sigma_{HeI}$ for \HI and \HeI.
The parameter $\epsilon_x$ sets the relative
contributions of the X-ray and soft UV component to the background
radiation field.  We consider four X-ray normalizations,
$\epsilon_x = 0$, $0.1$, $1$, and $10$, where $\epsilon_x=0$ denotes cases
with only the soft Lyman-Werner flux.

\begin{figure*}
\includegraphics{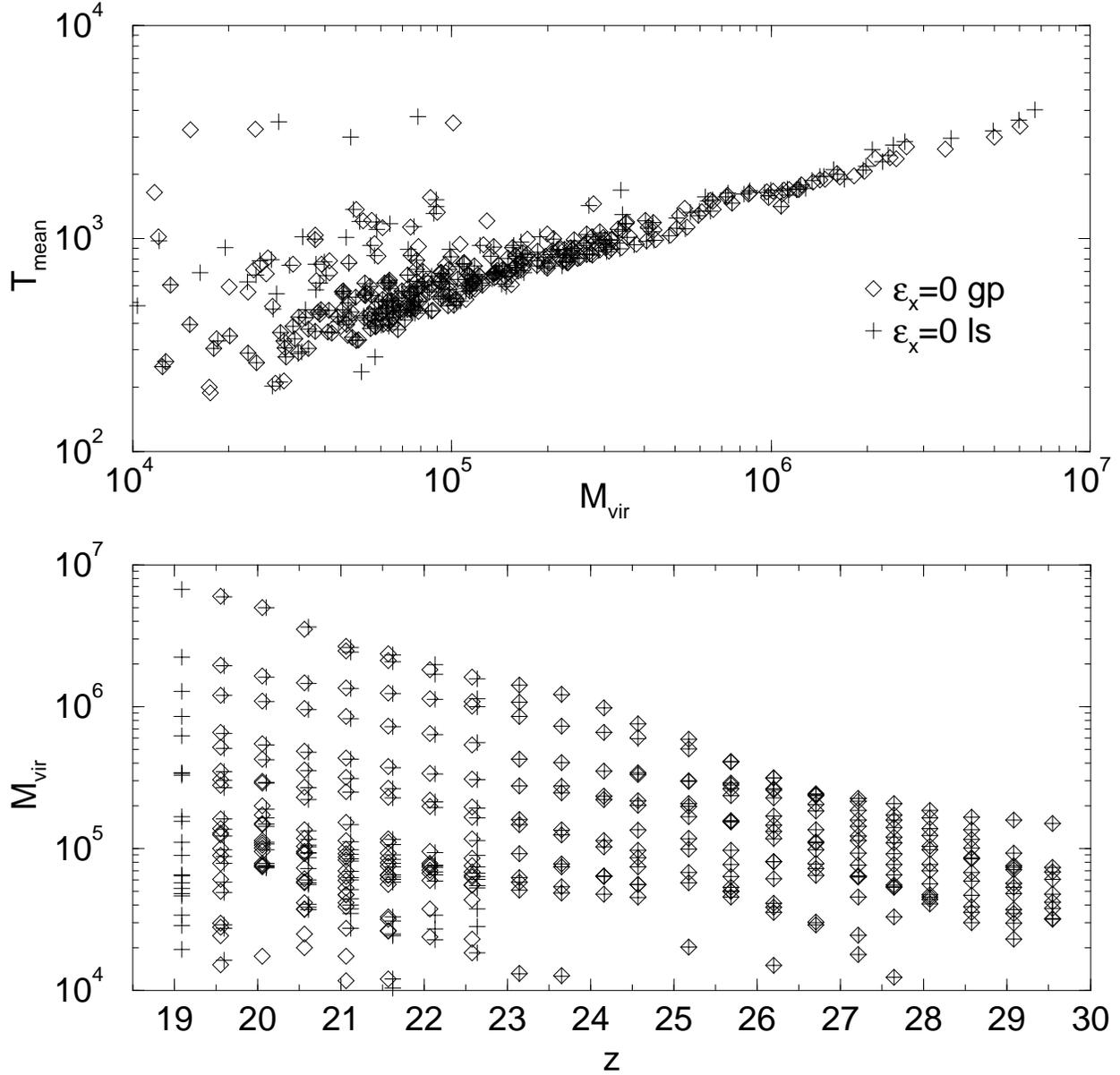}
\caption{(upper) Mean temperature as a function of virial mass for
pregalactic clouds in the presence of a Lyman-Werner flux
$F_{LW}=10^{-21}$~\fluxunit. (lower) Virial mass as a function of redshift
$z$ for pregalactic clouds in the presence of a Lyman-Werner flux
$F_{LW}=10^{-21}$~\fluxunit. Diamonds
(pluses) represent halos simulated using Galli \& Palla (Lepp \& Shull)
cooling functions.}
\label{fig:coolcomp}
\end{figure*}

Our simulations follow the nonequilibrium, time-dependent
evolution of nine chemical species (H, H$^+$, He, He$^+$, He$^{++}$,
e$^-$, \HH, H$^+_2$, H$^-$) using the algorithm of Anninos, \etal
(1997) initialized with post-recombination abundances (Anninos \&
Norman 1996). The rate coefficients for the reaction network for these species
are primarily those described in MBAI. We have in this work, however, updated
the \HH cooling function to that given in Galli \& Palla (1998). As a check
on how this affects our results, we simulate the $\epsilon_x=0$ (soft UV flux
only) case using both the Lepp \& Shull (1983) cooling function from previous
work and the Galli \& Palla (1998) cooling function used for the X-ray
simulations here. As shown in Figure \ref{fig:coolcomp},
there is statistically little difference between the characteristics of
the population of pregalactic objects produced using the Galli \& Palla (1998)
\HH cooling function in the presence of a Lyman-Werner UV flux
$F_{LW}=10^{-21}$~\fluxunit and those produced in an identical simulation
using the cooling function of Lepp \& Shull (1983). Cooling is somewhat more
efficient using the Galli \& Palla cooling function in the most massive peaks
($M \ga 2 \times 10^{6} \Ms$) in our sample, resulting in a slight drop in
mean temperature for these clouds and a modest increase in the fraction of
gas that can cool. (See  also Table \ref{tab:meanprops} and
\S\ref{sec:fractions}).
Although the dependence is weak, we present both of these cases,
$\epsilon_x=0$~gp using the Galli \& Palla cooling function and
$\epsilon_x=0$~ls using the Lepp \& Shull cooling function,
as a measure of the sensitivity of our
results to the parameterization of the cooling chemistry. A sixth
simulation ($\epsilon_x=0$,$F_{LW}=0$) evolves the same $z=30$ initial
conditions forward using Galli \& Palla cooling functions in the absence of
any external radiation field to test whether the X-rays can completely
overcome the effects of the photodissociating flux turning a negative
feedback effect into a net positive one.

The dominant reaction chain for the production of the \HH coolant,
\begin{equation}
H + e^- \rightarrow H^- + \gamma
\label{eq:hminusform}
\end{equation}
\begin{equation}
H + H^- \rightarrow H_2 + e^- ,
\label{eq:hminus2h2}
\end{equation}
is critically dependent on the electron abundance in the cloud.
When X-rays are present, the electron fraction increases both because of the
primary ionization of \HI, \HeI, and \HeII by the X-ray photon and because of
the production of secondary electrons through subsequent interactions of the
primary high energy electron with the gas. We follow the time evolution of
the electron fraction including the effects of the X-ray secondary electrons
on subsequent ionizations and heating of the gas.  
In general the fractions of the primary electron energy deposited as 
secondary ionizations and as heat depend on the energy of the primary electron.
However, for primary electron energies much greater than $100$~eV they 
rapidly approach a constant locus. Since most of the X-ray energy input 
comes from high energy photons, we use the analytic fits of Shull \& 
Van Steenberg (1985) to these asymptotic forms. 
Thus the fractions of primary
electron energy, $f_1$ and $f_2$,
available for further ionizations of \HI and \HeI are
\begin{equation}
f_1 = 0.3908 (1-x^{0.4092})^{1.7592}
\label{eq:energyfracH}
\end{equation}
for \HI and
\begin{equation}
f_2 = 0.0554(1-x^{0.4614})^{1.6660}
\label{eq:energyfracHe}
\end{equation}
for \HeI where $x$ is the ionization fraction, $x=n_{HII}/(n_{HI} + n_{HII})$.
The rate coefficients for HI and HeI ionization, including the effects of the
secondary electrons, are then given by
\begin{equation}
\begin{split}
k_{24}
= & \int_{\nu_{HI}}^\infty\frac{F_x\sigma_{HI}}{h_P\nu}\,d\nu 
+ f_1 \biggl
(\int^\infty_{\nu_{HI}}\frac{F_x\sigma_{HI}}{h_P\nu}\bigl
(\frac{\nu-\nu_{HI}}{\nu_{HI}}\bigr ) \,d\nu \\
 & + 
\frac{n_{HeI}}{n_{HI}}\int^\infty_{\nu_{HeI}}\frac{F_x\sigma_{HeI}}{h_P\nu}\bigl ( \frac{\nu - \nu_{HeI}}{\nu_{HI}}\bigr )\,d\nu \biggr )
\end{split}
\label{eq:ionHI}
\end{equation}
for \HI ionization and
\begin{equation}
\begin{split}
k_{26}
&= \int_{\nu_{HeI}}^\infty\frac{F_x\sigma_{HeI}}{h_P\nu}\,d\nu 
+ f_2 \biggl
(\frac{n_{HI}}{n_{HeI}}\int^\infty_{\nu_{HI}}\frac{F_x\sigma_{HI}}{h_P\nu}\bigl
(\frac{\nu-\nu_{HI}}{\nu_{HeI}}\bigr ) \,d\nu \\
& + 
\int^\infty_{\nu_{HeI}}\frac{F_x\sigma_{HeI}}{h_P\nu}\bigl ( \frac{\nu - \nu_{HeI}}{\nu_{HeI}}\bigr )\,d\nu \biggr )
\end{split}
\label{eq:ionHeI}
\end{equation}
for \HeI ionization with $\sigma_{HI}$ ($\nu_{HI}$) and
$\sigma_{HeI}$($\nu_{HeI}$) the ionization cross sections
(threshold frequencies) for \HI and \HeI, $n_{HI}$ and $n_{HeI}$ the \HI and
\HeI number densities, $F_x$ the ionizing X-ray radiation field,
$h_P$ Planck's constant, and $h_P\nu$ the ionizing photon's energy.
In practice the contribution of \HeII to the above processes is negligible
and so we drop it.
The fraction $f_3$ of the primary electron energy deposited as heat in the
gas is (Shull \& Van Steenberg, 1985)
\begin{equation}
f_3 =0.9971 (1-(1-x^{0.2663})^{1.3163}).
\label{eq:heatfrac}
\end{equation}

We self-consistently evolve the ionization fraction $x$ throughout the
simulation using it to evaluate the $x$-dependent rate and heating functions
(Equations \ref{eq:ionHI} - \ref{eq:heatfrac}) at
each time step after turn-on of the ionizing background and follow the
subsequent formation of structure in the simulation volume to redshifts
$z \sim 19.5$.
The dominant contribution to the rate equation for the electron density
from Equations \ref{eq:ionHI} and \ref{eq:ionHeI} comes from the secondary
ionizations of \HI produced by \HeI photoionized electrons.
As pointed out by Venkatesan, Giroux \& Shull (2001), our use of the
two-power-law extrapolation of $\sigma_{HeI}$, as in Abel \etal (1997) and
Haiman, Abel \& Rees (2000), overestimates the contribution of the high
energy end of the X-ray spectrum to \HeI ionization.  However because it
enters into both Equations \ref{eq:spectrum} and \ref{eq:ionHeI}, changes
in the cross-section tend to cancel out.  Indeed, adopting the
cross-section from Verner et al. (1996) increases the ionization rate by
only 35\%.  Since, as we will show, the primary effect is from ionization,
this amounts to using a value of $\epsilon_x$ which is 35\% larger than
quoted (the heating rate is also decreased by 50\%, which is only
important for the highest value of $\epsilon_x$).

In summary our data set consists of six simulations each identically
initialized at $z=30$, before the onset of significant cooling, and then
evolved self-consistently within the \LCDM cosmology to redshifts
$z \sim 19.5$. The simulations differ only by the type and normalization of
radiation fields present. Two simulations ($\epsilon_x=0$~gp and
$\epsilon_x=0$~ls) include only a soft
\HH photodissociating flux  ($F_{LW}=10^{-21}$~\fluxunit) and quantify the
sensitivity of our results to the  Galli \& Palla (1998) parameterization of
the \HH cooling function used here
compared to the Lepp \& Shull (1984) parameterization used in previous work.
Three of the simulations distinguished by their X-ray flux normalizations
$\epsilon_x=0.1$, $1$, and $10$, introduce an additional ionizing X-ray
background at $z=30$ to probe the effect of an early population of
miniquasars on subsequent structure formation. This is the main emphasis
of this work. For comparison, the sixth simulation
($\epsilon_x=0$,$F_{LW}=0$) evolves
structure in the region in the absence of any radiation field.
\section{Sample Characteristics}
\label{sec:data}

The locations of high-density halos
are identified using the HOP algorithm (Eisenstein and Hut
1998) acting on the dark matter density distribution.  We choose all dark
matter concentrations identified by the HOP algorithm to have dark matter mass
$m_{dm} > 4.0\times 10^{4}\,\Ms$ and an average dark matter particle mass
$m_{part} \leq 39\,\Ms$. 
Our results are not sensitive to the halo mass cutoff adopted above since 
objects near and below this cutoff ($M_{vir} \la {\rm a\,few} \times 
10^4 \Ms$) simply can not cool. The constraint on the average dark 
matter particle mass guarantees that our sample is
restricted to the region of highest dark matter mass resolution and that the
peaks selected for further analysis have not been significantly contaminated
by more massive dark matter particle elements that could have migrated into
the region from a lower resolution grid patch.
The mean properties and radial profiles of the
$\sim 300$ pregalactic objects selected from each simulation are then
calculated. As in MBAI, we define the virial radius $r_{vir}$ of the object
to be that radius of a sphere within which the average density of
the cloud exceeds $200$ times the mean density of the universe, the
virial mass $M_{vir}$ to be the total mass enclosed within that radius, and
the virial temperature $T_{vir}$ (given by Equations 5 and 6 in MBAI) to be
the expected mean gas-mass-weighted temperature of a cloud in the absence of
any external heating and before cooling starts.

\begin{figure*}
\includegraphics{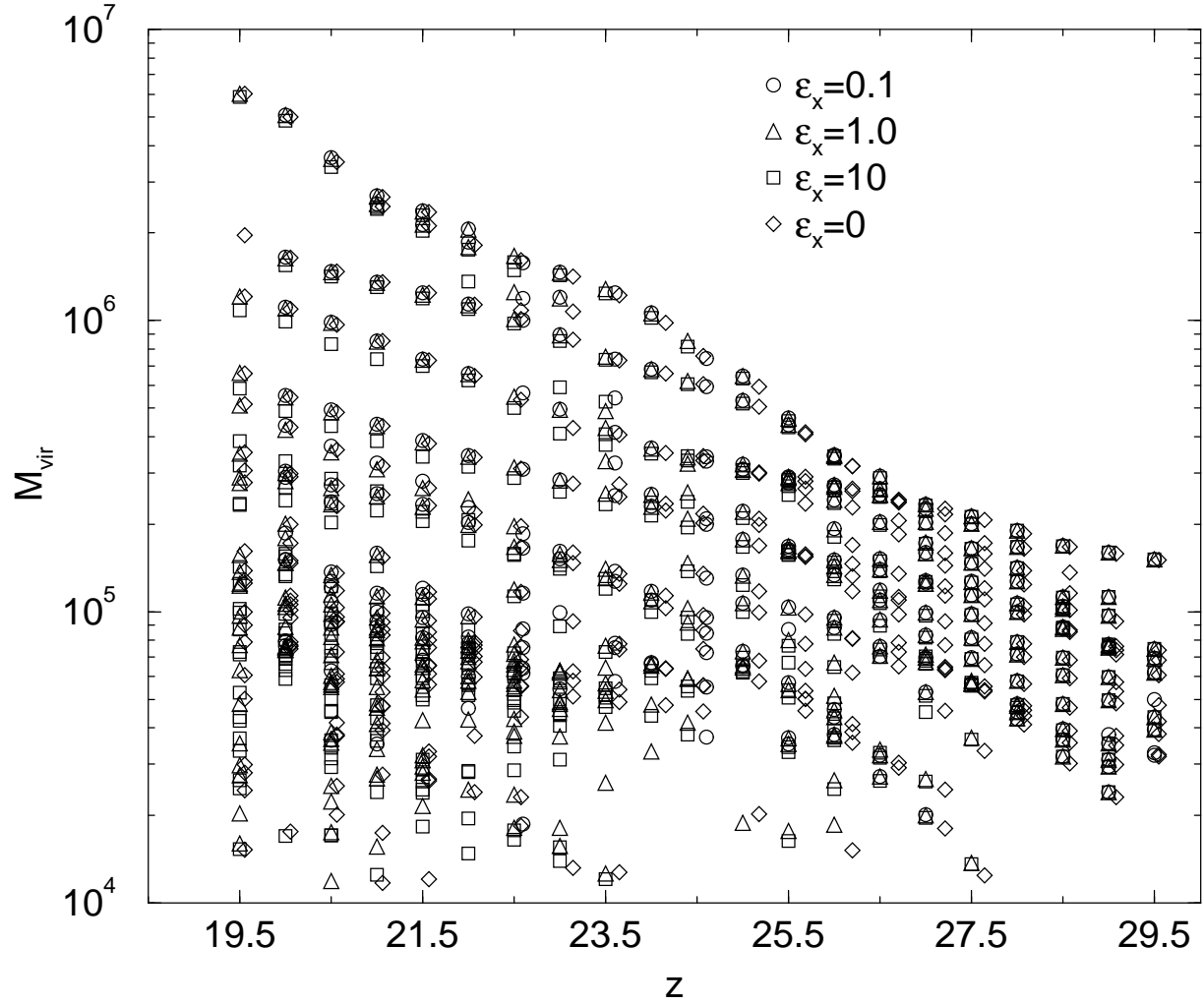}
\caption{Virial mass as a function of redshift for peaks used in the analysis
sample with Lyman-Werner flux $F_{LW}=10^{-21}$~\fluxunit and X-ray flux
normalizations of $\epsilon_x=0$ gp (diamonds), $0.1$ (circles), $1$
(triangles) and $10$ (squares).}
\label{fig:Mvirvz}
\end{figure*}

In Figure \ref{fig:Mvirvz} we show the masses of the pregalactic objects in
our sample as a function of redshift for the simulations with
$\epsilon_x=0$ gp, $0.1$, $1$, and $10$.
The horizontal shift seen in the $\epsilon_x=0$ gp and $\epsilon_x=0.1$ data is
due to our choice of output redshifts for those simulations. The
assignment of virial masses below the dark matter selection limit of
$4 \times 10^4 \Ms$ for a few of the peaks in our sample is caused by
differences in the definition of a peak's extent in the HOP (peak
identification) algorithm versus $r_{vir}$ in the spherically averaged radial
profile analysis.
Since these low mass objects can not cool, they do not affect our subsequent 
results in any way.
We see in Figure \ref{fig:Mvirvz} the characteristic pattern of
hierarchical structure formation whereby larger objects form from smaller
ones through merging. Although mergers are apparent, especially in the higher
mass peaks, we do not follow the evolution of all the low-mass clouds
in our simulation and, therefore, do not construct complete merger trees for
objects in the data sample. Thus, as in MBAI, we
caution the reader that all of the peaks are not statistically independent.
In particular, this results in reduced scatter in the measured properties
for the highest mass peaks, which are rare in our simulation volume.

When we compare the results of our several simulations, we find that the
redshift distributions of virial masses for peaks in our data are
very similar, nearly independent of the level of X-ray flux. This is to be
expected since the gravitational potential controlling the formation of the
cloud is dominated by dark matter everywhere except for the innermost core
after collapse and the dark matter is unaffected by the internal
cooling chemistry. There is, however, a small, systematic shift to lower
virial masses seen at redshifts $z \la 24$ for clouds exposed to the
highest level of X-ray flux ($\epsilon_x=10$). This is consistent with X-ray
heating of gas within those clouds making it more difficult for the gas
to be trapped in their dark-matter-dominated gravitational potentials.

\begin{figure*}
\includegraphics{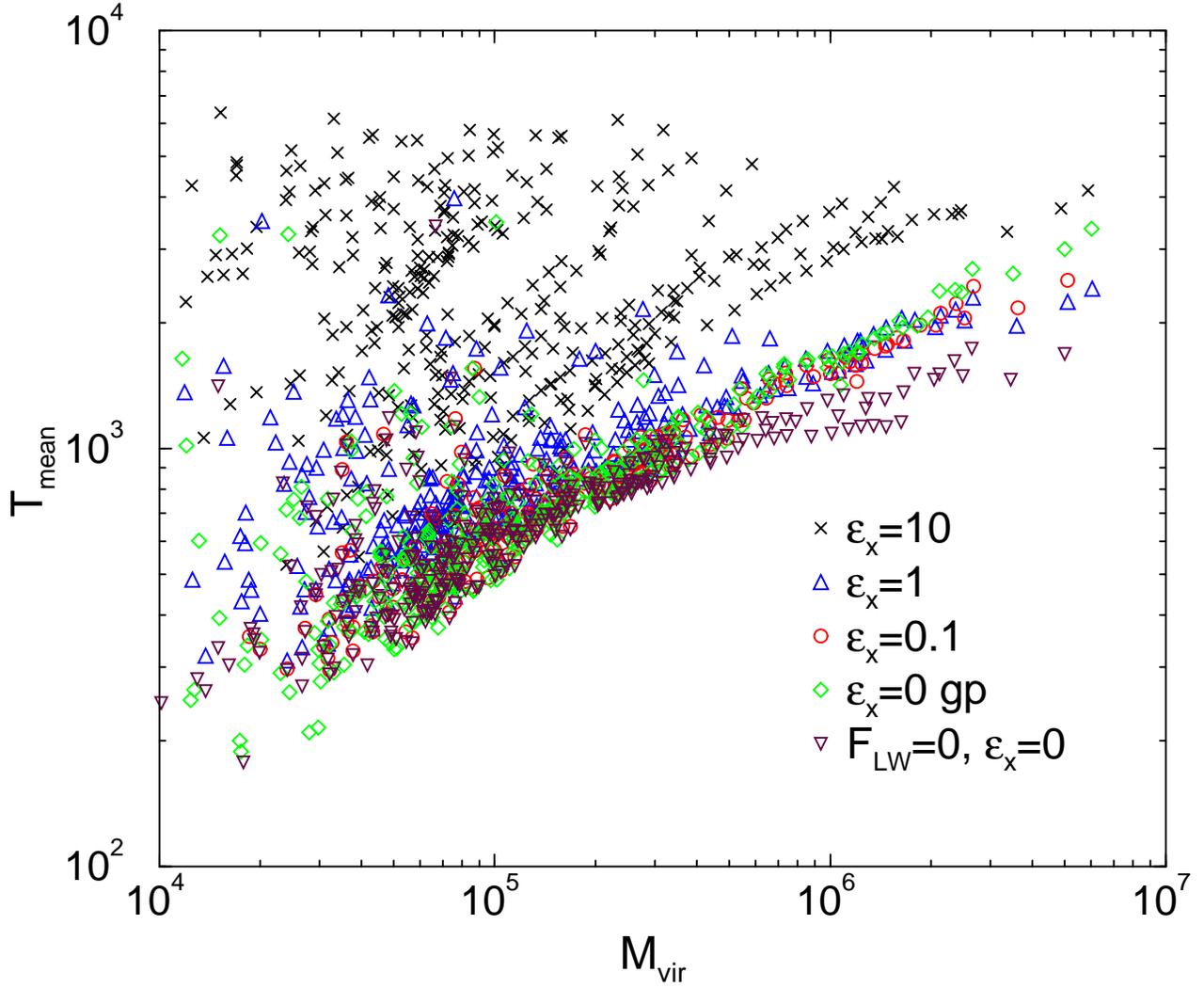}
\caption{Mean gas-mass-weighted temperature as a function of virial mass for
pregalactic clouds in the presence of a Lyman-Werner flux
$F_{LW}=10^{-21}$~\fluxunit with no X-ray background ($\epsilon_x=0$ gp,
green diamonds) and with a
soft X-ray background whose relative normalizations are
$\epsilon_x = 0.1$ (red circles), 1 (blue triangles up) and 10 (black crosses).
The case with no external radiation fields ($F_{LW}=0$, $\epsilon_x=0$) is
represented by maroon down-pointing triangles.  While most points fall along
the virial relation, the $\epsilon_x = 10$ case shows substantial
heating from X-rays.  In addition, cooling becomes important for the
most massive objects.}
\label{fig:TmeanMvir}
\end{figure*}

\begin{figure*}
\includegraphics{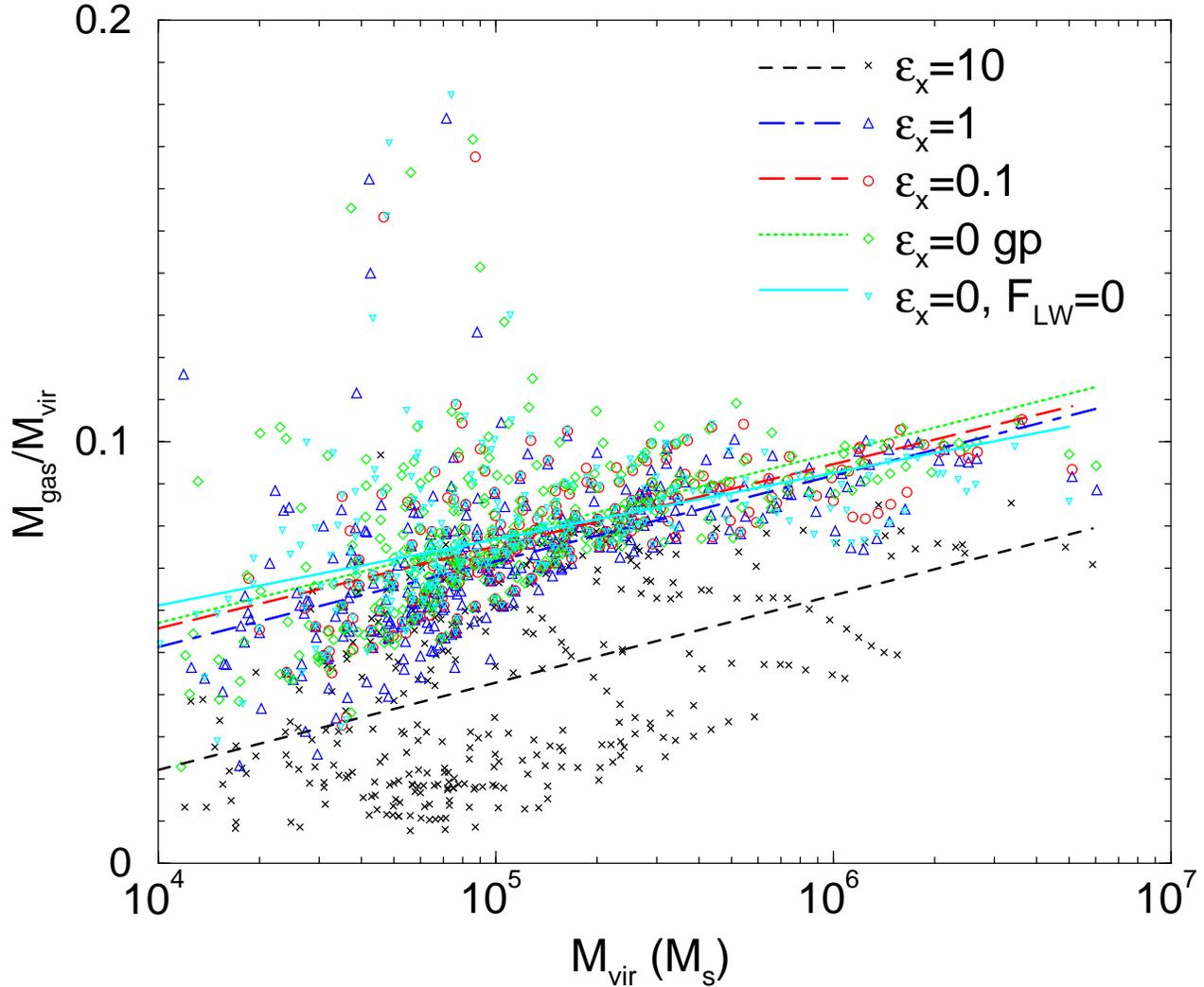}
\caption{Gas fraction as a function of virial mass for clouds used in the
analysis sample with Lyman-Werner flux $F_{LW}=10^{-21}$~\fluxunit and
X-ray flux normalizations $\epsilon_x=0$~gp (green diamonds), $0.1$
(red circles), $1$ (blue triangles up), $10$ (black crosses). Maroon 
triangles down represent the $F_{LW}=0$, $\epsilon_x=0$ no radiation field 
case. Lines represent
a mean regression analysis of the data set with respect to $\ln(M_{vir})$
for cases $\epsilon_x=0$~gp (green dotted), $0.1$(red long dashed), $1$(blue 
dot dashed), $10$(black dashed), and no radiation fields (maroon solid).  
While most objects fall near to the global ratio of 0.08, the 
$\epsilon_x = 10$ case shows the effect of X-ray heating.}
\label{fig:MgasMtot}
\end{figure*}

This X-ray heating of low mass clouds is  more apparent in
Figure \ref{fig:TmeanMvir} where we show
the mean gas-mass-weighted temperature as a
function  of virial mass
for pregalactic clouds exposed to a mean
Lyman-Werner flux $F_{LW}=10^{-21}$ \fluxunit and X-ray
flux varying over two orders of magnitude. We also show the $\epsilon_x=0$,
$F_{LW}=0$ case (no background radiation fields) for comparison.
For $\epsilon_x =1$ there is a
modest increase in the mean temperature of the cloud for masses
$M_{vir} \la 5 \times 10^5 \Ms$;
while for $\epsilon_x=10$ heating is
dramatic with the mean gas temperature of the clouds raised well above their
virial temperatures for $M_{vir} \la 10^6 \Ms$.
In Figure \ref{fig:MgasMtot} we investigate the consequences of this heating
by plotting the cloud gas fraction $M_{gas}/M_{vir}$
as a function of $M_{vir}$ for halos in our data sample. The lines are from
mean regression analyses of the data for each level of X-ray flux. Because
there is large scatter in the data (correlation coefficients of only
$\sim 0.5$ -- $0.6$ ) the lines are only meant to guide the eye. However the
decrease of gas fraction with increasing X-ray flux is clear with mean gas
fractions given by $\sim 0.08$, $0.08$, $0.07$, and $0.04$ for X-ray flux
normalizations $\epsilon_x = 0$~gp, $0.1$, $1$, and $10$. The mean gas
fractions for the low flux cases ($\epsilon_x \leq 1$) agree with
the mean gas fraction ($0.08$) in the sample of halos not exposed to an
external radiation field.  This is consistent with most of the accreted gas
being retained by the gravitational potential. However,
heating in clouds experiencing the highest X-ray flux, such
as those nearby to a newly formed miniquasar, causes a significant
fraction of the gas to be evaporated into the surrounding intergalactic medium.
Such effects have interesting consequences for re-ionization (Haiman,
Abel \& Madau 2001).

In Figure \ref{fig:TmeanMvir}
we also see that X-ray enhanced cooling (positive feedback) occurs
in the most massive peak in the simulation once its mass
exceeds $2 \times 10^6 \Ms$.
Above this mass the mean temperature of the
cloud exposed to both X-rays and the Lyman-Werner flux is
noticeably lower than for the case with the Lyman-Werner flux alone and the
mean temperature for a given mass decreases with increasing X-ray flux for
$\epsilon_x \leq 1$. However, in all cases the mean temperature of the cloud
lies above the limiting case with no radiation fields present. Furthermore
the onset of cooling in the absence of background radiation fields occurs
at $M_{vir} \sim 2 - 3 \times 10^5 \Ms$, nearly an order of magnitude sooner.
For the maximum X-ray flux level we consider, $\epsilon_x=10$, the situation
is even worse. While cooling is apparent in the high mass objects
($M_{vir} \ga 10^6 \Ms$) it does not overcome the effects of
X-ray heating and drop to its virial temperature until
$M_{vir} > 2 \times 10^6 \Ms$. The mean temperature remains above that
found for clouds exposed to the \HH photodissociating flux alone
($\epsilon_x=0$~gp) up to the highest masses we find in our sample.
Thus the presence of the X-ray background, even at these high levels, is
insufficient to turn the negative feedback effect of the softer
\HH photodissociating UV background into a net positive one.

In Table \ref{tab:meanprops} we present a quantitative example by
comparing the mean masses, gas fractions, and
temperatures found in our simulations for the most massive pregalactic
cloud at redshift $z=20$ when it
has grown through merging to a mass $M_{vir} \approx 5 \times 10^6\,\Ms$
($T_{vir} \approx 5300$~\kel) and radius $r_{vir} \approx 239$~pc.
First in the absence of any X-ray flux we see that use of
Galli \& Palla (1998) cooling functions does
increase cooling in this cloud producing about a $10 \%$ reduction in its
mean temperature over that obtained using cooling functions by
Lepp \& Shull (1983). Second, positive
feedback does occur, but the effect is quite modest. Exposing
the cloud to increasing levels of X-rays initially promotes cooling, as
expected, causing the mean cloud temperature to decrease by $16 \%$ for
$\epsilon_x=0.1$ and $25 \%$ for $\epsilon_x=1$ over the case with no X-rays
and Galli \& Palla cooling. However even in the maximal $\epsilon_x=1$ case,
the mean temperature is $33 \%$ higher in this cloud than in the cloud evolved
from the same $z=30$ initial conditions but with no radiative feedback.
Finally once the X-ray background becomes
sufficiently strong, heating outside the core dominates. For $\epsilon_x=10$
the mean temperature of this cloud increased by $25 \%$ over that of the
cloud  exposed to only the soft Lyman-Werner UV background and is more than
a factor of $2$ above the mean temperature found for the cloud without
radiative feedback. Furthermore, the fraction of gas retained by the
gravitational potential decreased from
$9 - 10 \%$ of the cloud mass for $0 \leq \epsilon_x \leq 1$ to
$7 \%$ for $\epsilon_x=10$, another reflection of the evaporation of X-ray
heated gas from the outer radii in even this most massive cloud in our
data set.
\section{Cold Gas Fractions}
\label{sec:fractions}

A critical input parameter into semi-analytical structure formation
and reionization models that include stellar feedback is the amount of gas in
pregalactic objects that is available to form stars.  In MBAI we used our
data sample to determine both the fraction of gas in these low mass
pregalactic objects that can cool $f_c$ and the fraction of gas $f_{cd}$
that can both cool and become dense in the presence of a mean Lyman-Werner
\HH photodissociating flux, thus quantifying the negative feedback effect of
the soft UV radiation produced by the first stars on subsequent
star formation. In this section we address how these gas fractions,
$f_c$ and $f_{cd}$, change when an ionizing X-ray background is present and
whether at some level they can reverse, as suggested by Haiman, Abel \& Rees
(2000), the negative feedback of the \HH photodissociating flux.

As in MBAI we define $f_c$ and $f_{cd}$ in the following way:
\begin{quote}
$f_c$ is the fraction of total gas within the virial radius with
temperature $T < 0.5 T_{vir}$ and gas density $\rho > 1000 \rho_{mean}$
where $\rho_{mean}$ is the mean gas density of the universe. This is the
amount of gas within the cloud that has been able to cool due to molecular
hydrogen cooling.

$f_{cd}$ is the fraction of total gas within the virial radius with
temperature $T < 0.5 T_{vir}$ and gas density $\rho > 10^{19} \MsMpc \sim
330$ \cmc .  This is the fraction of gas within the cloud that is available
for star formation.
\end{quote}
The above criteria are applied on a cell by cell basis within the virial
radius for each peak in the data sample described in \S\ref{sec:data}. The
temperature threshold has been chosen to ensure that the gas is substantially
cooler than the virial temperature of the peak. The density
threshold for cold, dense gas corresponds to the gas density at which the
baryons become important to the gravitational potential and thus to the
subsequent evolution of the core. The density threshold for the cooled gas
is chosen to minimize the contribution of cold, infalling substructure
and of cool gas within $r_{vir}$ but still outside the virial shock.
(Please see MBAI Section 4 for more detail.)

\begin{figure*}
\includegraphics{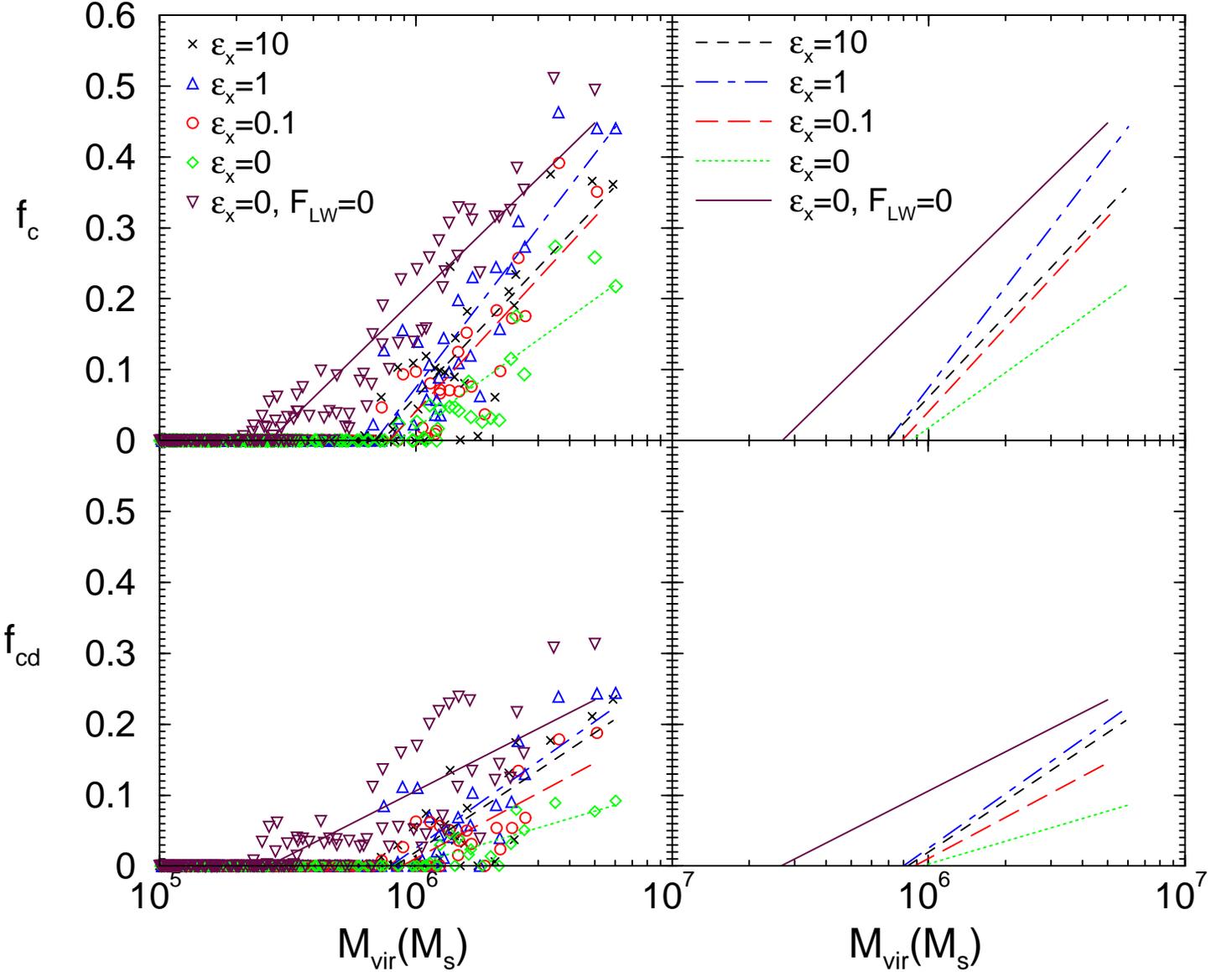}
\caption{Fraction of cold gas within the virial radius as a
function of cloud mass and X-ray flux for a soft Lyman-Werner UV background
flux $F_{LW}=10^{-21}$~\fluxunit. In the top panels, we plot $f_c$,
the fraction of gas that has cooled via \HH cooling
($T < 0.5 T_{vir}$, $\rho > 1000 \rho_{mean}$) with $\rho_{mean}$ the mean
baryonic density of the universe, while in the lower panels, we show
$f_{cd}$, the fraction of cold, dense gas ($T < 0.5 T_{vir}$,
$\rho > 10^{19}$~$\Ms$~Mpc$^{-3}$) available for star formation. Maroon
triangles down show the limiting $F_{LW}=0$, $\epsilon=0$ case with no
radiation fields present. Lines represent mean regression analyses of
$f$ with the logarithm of the cloud mass for the cases 
$\epsilon_x=0$~gp (green dotted), $0.1$
(red long dashed), $1$ (blue dot dashed), $10$ (black dashed), and $F_{LW}=0$, 
$\epsilon_x=0$(maroon solid).  The right-hand panels show the same fitted
lines as in the left-hand panels, but plotted without data for clarity.
}
\label{fig:coldfractionstats}
\end{figure*}

In Figure \ref{fig:coldfractionstats}
we show the fraction of gas that can cool (top panel) and the fraction of gas
that can both cool and become dense (bottom panel) and thus be available
for star formation for each of our
simulations ($\epsilon_x=0$~gp, $0$~ls, $0.1$, $1$, $10$ and
$F_{LW}=0$ $\epsilon_x=0$). In Table
\ref{tab:coldfracfits} we show the results of a mean regression analysis of
$f_c$ and $f_{cd}$ with the logarithm of the cloud mass,
\begin{equation}
    f_i = B_i \ln (M/M_i)
\label{eq:regression}
\end{equation}
for $i=c$ (cooled gas), $i=cd$ (cold, dense gas) and $M > M_i$.
$M_i$ and $K_i$ ($i=c,cd$) are the mass thresholds and correlation
coefficients for each process derived from the regression analyses.
We again see that the fractions of gas that can cool ($K_c \sim 0.8$) or
cool and become dense ($K_{cd} \sim 0.9$) are highly correlated with the mass
of the pregalactic cloud, although as discussed in \S\ref{sec:data} the
reduced scatter in the high mass end is partially due to the small number of
independent high mass peaks in our simulation volume.

The most striking aspect of Figure \ref{fig:coldfractionstats}
is that the positive feedback of an X-ray background on structure formation
is remarkably weak, even though we vary the X-ray intensity by two orders of
magnitude. As the relative
normalization of the X-ray flux is increased to
$\epsilon_x=1$, the mass threshold at which gas can cool or cool and become
dense (and thus available for star formation)
does decrease as expected if the positive effect of X-rays on \HH formation
partially compensates for the \HH destruction by the soft UV Lyman-Werner
photons, but only modestly.
This decrease in the mass threshold for gas to cool,
from $\sim 8.5 \times 10^5\, \Ms$ for $\epsilon_x=0$ to
$\sim 7 \times 10^5\, \Ms$  for $\epsilon_x=1$,  and for gas to both cool and
become dense, from $10^6 \Ms$ for $\epsilon_x=0$ to
$8 \times 10^5 \Ms$ for $\epsilon_x=1$,
caused by the presence of the X-ray background
is much less than the decrease
from $10^6 \,\Ms$ to $4.2 \times 10^5\,\Ms$ in these thresholds found in
MBAI caused by
an order of magnitude reduction in the \HH photodissociating flux
(from $10^{-21}$ to $10^{-22}$ \fluxunit) and the mass thresholds for
the case with $\epsilon_x=1$ are still a factor $\ga 3$ higher
than for the case when no Lyman-Werner UV background is present.
Thus, in contrast to Haiman, Abel  \& Rees (2001), we find that the positive
effect of the X-rays is too weak to overcome the delay in cooling caused by
the \HH photodissociating flux.  As the X-ray
flux is increased further to $\epsilon_x=10$, the mass threshold for cooling
increases again, because at these X-ray
flux levels the weakly enhanced cooling must compete with significant X-ray
heating of the gas within the cloud. This is consistent with the fact that
the redshift when maximal refinement first occurs, i.e. when the first peak
``collapses'' in the simulation volume,
increases with increasing X-ray flux for $0 \le \epsilon \le 1$
($z=21.5$, $22.6$, $23.2$ for $\epsilon=0$~gp and ls, $0.1$, $1$ respectively)
, decreases slightly
($z=23$) for $\epsilon=10$, but in all cases occurs significantly
later than $z=26.5$, the redshift of maximal refinement in our no radiation
field control simulation.

The effect of the X-ray background on the amount of gas that can cool and
become dense once cooling commences is more significant, but still small.
In Figure
\ref{fig:coldfractionstats}
and Table \ref{tab:coldfracfits} we see that
the regression coefficients $B_i$, $i=c,cd$ increase by as much as a factor
$\sim 2$ with increasing
X-ray flux for $\epsilon_x \leq 1$. This trend
reverses for the highest flux level we consider ($\epsilon_x=10$) because
then X-ray heating of the gas within the cloud is important,
reducing the fraction of gas that has been able to cool to temperatures
significantly below the cloud's virial temperature to values similar
to that obtained with $\epsilon_x=0.1$, an X-ray flux two orders of
magnitude smaller. The slopes for the two most intense X-ray flux levels
($B_c = 0.17- 0.20$ and $B_{cd} \sim 0.1$) are also
somewhat steeper than those ($B_c = 0.15$ and $B_{cd} \sim 0.08$) found in
our $F_{LW}=0$, $\epsilon_x=0$ control simulation. Better
statistics are needed for high mass clouds to determine whether this
steepening is significant. Note, however, that in the most massive peak at
$z=20$ the fraction of gas available
for star formation in the $F_{LW}=0$ control simulation ($0.31$) is still
significantly higher than the maximum found ($0.24$) for that cloud when both
X-rays and the soft Lyman-Werner background are present. We also find that
use of the Galli \& Palla (1998) rather than Lepp \& Shull (1984) fit for the
\HH cooling function increases the slope of the
fitting function for the cold and dense gas available for star formation
by $\sim 30\%$. As in MBAI we attribute the softer slope for the
$\epsilon_x=0$~gp(ls) cases to poor statistics and increased scatter near the
cooling threshold.
\section{Internal Cloud Properties}
\label{sec:profiles}

\begin{figure*}
\includegraphics{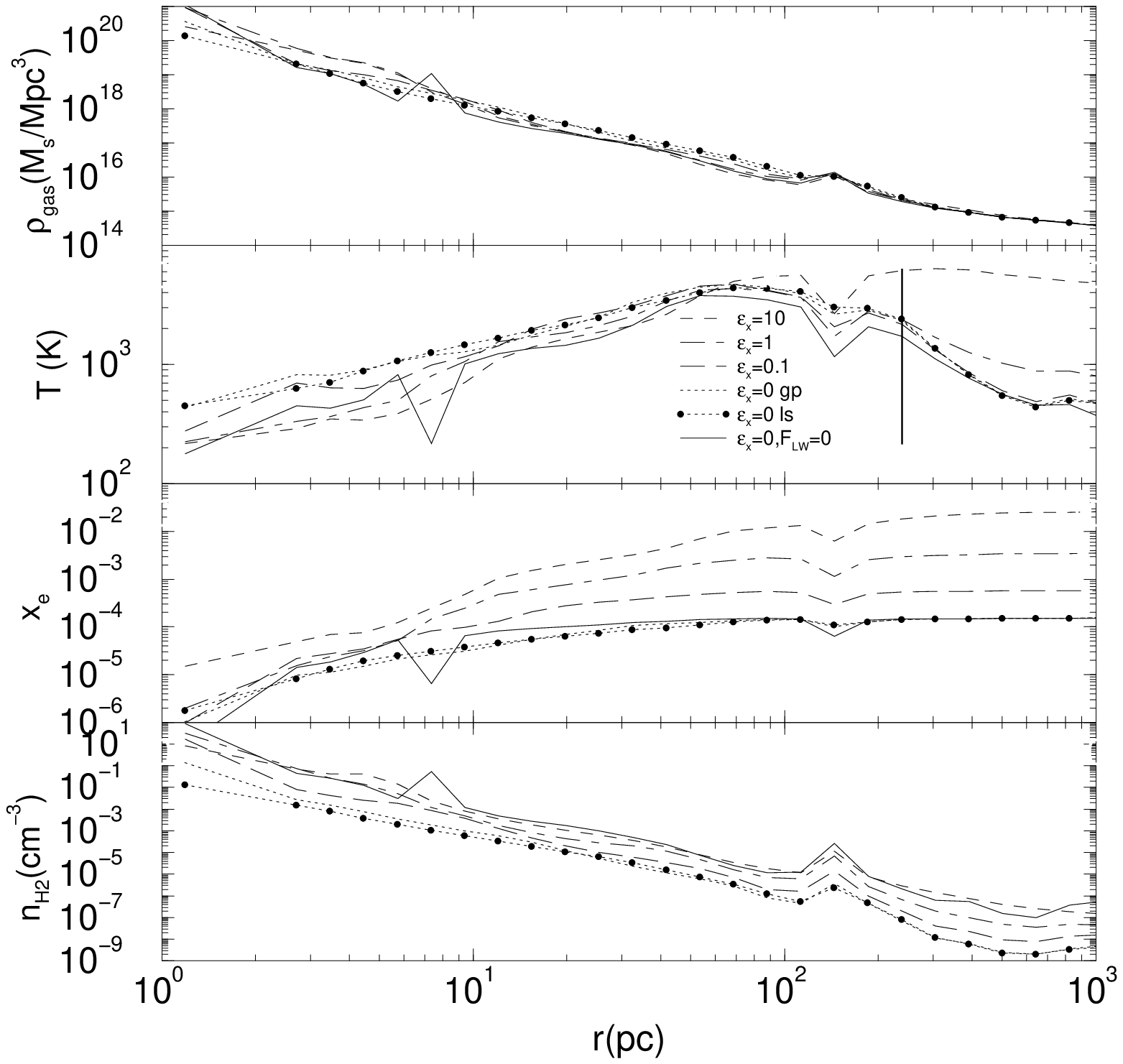}
\caption{Radial profiles of (from top to bottom) the gas density
$\rho_{gas}$, temperature $T$, electron abundance $x_e$ and
\HH number density $n_{H2}$ for
a pregalactic cloud of mass $5 \times 10^6 \Ms$ at $z=20$ with
$F_{LW}=10^{-21}$~\fluxunit and relative
X-ray flux normalizations $\epsilon_x=0$ gp (dotted), $0$ ls
(dotted with filled circles), $0.1$ (long dashed), $1$ (dot-dashed), and 
$10$ (dashed) and
with no background radiation fields (solid).
The vertical line denotes the virial radius $r_{vir}=239$~pc. }
\label{fig:profileprops}
\end{figure*}

We can gain a better understanding of the physical processes at work
when pregalactic structure is exposed to both X-ray and Lyman-Werner UV
backgrounds (and hope to explain why the effect is so mild) by studying the
internal structure of a collapsing cloud. In this section we compare radial
profiles of physical properties for a single cloud for the various levels of
X-ray and soft UV flux used in our simulations. Guided by
Figure \ref{fig:TmeanMvir}
and Table \ref{tab:meanprops}, we choose the
most massive peak in our simulations ($M_{vir} \sim 5 \times 10^6 \Ms$)
at $z=20$ for these comparison studies.
In Figure \ref{fig:profileprops} we present the spherically averaged radial
profiles of the gas density $\rho_{gas}$, temperature $T$, electron
abundance $x_e = n_e/(n_{HI}+n_{HII})$  and
\HH number density $n_{H2}$ for this cloud.
The virial radius $r_{vir}=239$~pc is denoted by a vertical line
in the gas temperature panel. The enhancement
seen in the gas density and \HH number density (corresponding to the dip in
temperature and electron abundance) for radii $100 < r < 200$~pc is the
characteristic signature of infalling substructure. As is often the case in
hierarchical models, the most massive peak lies within a dynamically active
filament in the simulation volume where mergers are common and, in fact,
has only recently formed ($z=20.5$) through the completion of the merger of
two nearly equal mass subcomponents (see the massive peaks in MBAI, Figure 5).

The most remarkable feature of the profiles in Figure \ref{fig:profileprops}
(and one of the main results of this paper) is how weakly the
X-rays affect the cloud properties, particularly
the gas density and temperature, even though we vary
the X-ray flux by two orders of magnitude. The gas density profiles at radii
$r > 10$~pc for the various X-ray fluxes are nearly indistinguishable,
although they do steepen somewhat for $\epsilon_x=1$ and $10$  and approach
more closely the density profile for the no radiation field case
($F_{LW}=0$, $\epsilon_x=0$).
The gas density profiles differ most in the inner $10$~pc of the structure.
The density tends to increase with increasing X-ray flux for fixed soft UV flux
$F_{LW}=10^{-21}$~\fluxunit. However, it appears to saturate for
$\epsilon_x=1$ with the profile flattening significantly for
$\epsilon_x = 10$, the highest X-ray
flux case.  While the maximum density shown for the $\epsilon_x=1$ simulation
is similar to that for the case with no
radiation field, substructure is much more obvious in the latter. This is
a consequence of the lower mass threshold (See \S\ref{sec:fractions})
for a cloud to become dense and collapse.
Thus infalling substructure is more highly evolved (colder and
denser) when no background fields are present.

The temperature panel of Figure \ref{fig:profileprops} is particularly
interesting. We find that the dominant effect of the higher two X-ray flux
levels is to heat and (from the third panel) partially ionize the lower
density gas near the virial radius. This serves to weaken the accretion shock
as the X-ray flux is increased until for our extreme case
($\epsilon_x=10$), the temperature of
the lower density IGM exceeds the virial temperature of the cloud.  Interior
to the accretion shock and over most of the object's extent
($10 \la r \la 200$~pc), the temperature  profiles are remarkably
similar. The onset of cooling within the cloud occurs near $r \sim 60$~pc
for all the simulations where the density has reached
$\rho \sim 3 \times 10^{16}$~Mpc$^{-3}$
($n \sim 1$~\cmc), in agreement with Haiman, Abel \& Rees (2000). The
temperature does decrease modestly with increasing X-ray flux demonstrating
positive feedback.  As the X-ray flux is increased, thereby increasing the
electron fraction through secondary ionizations, there is more \HH coolant
produced so that the cloud cools more efficiently (See the lower two panels
of Figure \ref{fig:profileprops}).  Within $10$~pc
sufficient \HH has formed by $z=20$ to drive the temperature to its minimum
for molecular hydrogen alone, $\sim 200 \kel$, in the high X-ray flux
cases ($\epsilon_x=1$, $10$) and the no radiation field case
($F_{LW}=0$,$\epsilon_x=0$). The temperature is close to
its minimum for the other cases ($\epsilon_x=0.1$,$0$~ls, $0$~gp).
Again, the major difference between the profiles is that in the
$F_{LW}=0$, $\epsilon_x=0$ case infalling substructure at $r \sim 7.4$~pc
is clearly seen; while it is not seen in any of the profiles that include
Lyman-Werner and X-ray backgrounds, even those with the highest X-ray
fluxes (see MBAI for more discussion of substructure in halos, in
particular Figure 5 of that paper).
For the no radiation field case this substructure has formed sufficient
molecular hydrogen ($n_{H2} \sim 0.5$) to cool to
$\sim 200$~\kel (the minimum temperature for \HH cooling)
and reach high density; while in all other cases the cooling
has been delayed within the substructure by the \HH photodissociating soft
UV component of the background radiation field.

\begin{figure*}
\includegraphics{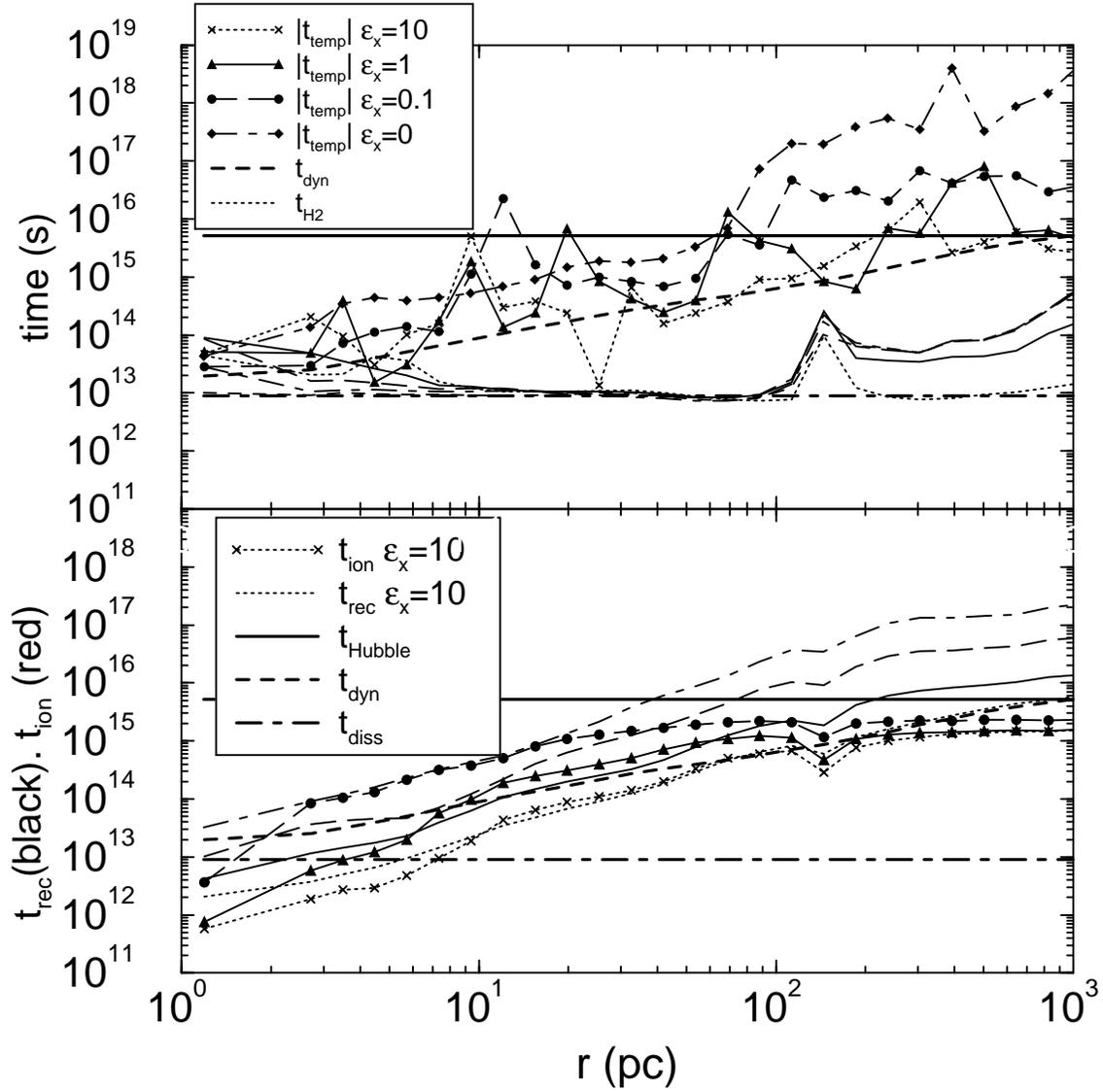}
\caption{Radial profiles of timescales relevant to cooling and collapse at
z=20 for the same cloud as in Figure \protect\ref{fig:profileprops}. 
In the top panel we plot $t_{temp}$ as lines with points for the four
values of $\epsilon_x$.  These times are generally longer or equal to
the dynamical time (shown as a dashed line for just one model because
all four cases are quite similar).  On the other hand, the \HH
formation times for the four cases (same line styles as for $t_{temp}$
but without the points are very close to the \HH photodissociation
time which is ploted as a dot-dashed line.  In the bottom panel, we
compare the ionization and recombination times for the four values of
$\epsilon_x$ using the same line convention as in the top panel
($t_{ion}$ is plotted with symbols, while $t_{rec}$ is not).  Note
that one cannot define $t_{ion}$ for the $\epsilon_x=0$ case so we 
do not plot it.  The two
time-scales are comparable either for high densities (i.e. small
radii) or large values of the X-ray flux, $\epsilon_x$.
}
\label{fig:comparetimes}
\end{figure*}

In Figure \ref{fig:comparetimes} we plot timescales important to cloud
cooling and collapse for the same pregalactic cloud, redshift, and
X-ray background levels used in Figure \ref{fig:profileprops}.
In the top panel we show the \HH formation time
$t_{H2}$, \HH photodissociation time $t_{diss}$, and the magnitude of the
timescale for temperature change (loosely, the ``cooling'' time)
$|t_{temp}|$  compared to the dynamical time $t_{dyn}$ and Hubble 
time $t_{Hubble}$. In the lower panel we compare the
recombination time $t_{rec}$ and ionization time $t_{ion}$ to 
$t_{Hubble}$ and $t_{diss}$.
All of the timescales except $t_{temp}$ and $t_{ion}$ are given by the
equations in MBAI. We repeat them here for completeness:
\begin{equation}
t_{dyn} =  \biggl(\frac{3\pi}{16G \rho}\biggr)^{1/2}, \notag
\end{equation}
\begin{equation}
t_{Hubble} =  \frac{2}{3H_0} (\Omega_0 (1+z)^3)^{-1/2}, \notag
\end{equation}
\begin{equation}
t_{H2} = \frac{n_{H2}}{k_7 n_{HI} n_e},
\end{equation}
\begin{equation}
t_{diss}  = k_{diss}^{-1} = 9 \times 10^{-9}/F_{LW}, \notag
\end{equation}
\begin{equation}
t_{rec} = 7.7 \times 10^9 T^{0.64}/n_e, \notag
\notag \end{equation}
where $G$ is the gravitational constant, $\rho$ is the total density,
$H_0$ is the Hubble constant today, $\Omega_0$ is the fraction of the
critical density carried in matter today, $z$ is the redshift, $n_{H2}$,
$n_{HI}$, and
$n_e$ are the molecular hydrogen, neutral hydrogen, and electron number
densities, $F_{LW}$ is the mean Lyman-Werner UV flux, $T$ is the temperature
and $k_7 \approx 1.8 \times 10^{-18}T^{0.88}$~cm$^3$\,s$^{-1}$ over this
temperature range.

We have defined a timescale for temperature change due to radiative
processes (very loosely, a generalization of the ``cooling'' time)
that includes both radiative cooling and the effects of heating by
X-ray photo-electrons
\begin{equation}
t_{temp} = \frac{e}{de/dt|_r}
\label{eq:tcool}
\end{equation}
where $e$ is the thermal energy of the gas and $de/dt|_r$ is its rate
of change due to radiative processes, such that $t_{temp} > 0$
indicates net cooling, while $t_{temp} < 0$ indicates net heating. The
sharp fluctuations in $|t_{temp}|$ seen in Figure
\ref{fig:comparetimes} when ionizing X-rays are present result from
the competition between regions where heating versus cooling dominates
within the radial shells.
Most often the net effect of the X-ray flux on the outer parts of the cloud
is to heat the gas; while net cooling is more likely in the interior regions.

The ionization time is given by
\begin{equation}
t_{ion}= \frac{n_e}{\epsilon_x n_{HI} R }
\label{eq:tion}
\end{equation}
where $n_e$, $n_{HI}$ and $n_{HeI}$ are the electron, neutral hydrogen and
neutral helium number densities, and $R = k_{24} + (n_{HeI}/n_{H})k_{26}$ is
the ionization rate in the gas (See Equations \ref{eq:ionHI} and
\ref{eq:ionHeI}). The ionization rate is dominated by the production of
secondary electrons from \HI produced by primaries from the X-ray
photoionization of  \HeI. Although we evolve the true ionization fraction
$x = n_{HII}/(n_{HI}+n_{HII})$ at each timestep in our calculation,
$n_{HII} \approx n_e$ in our simulations so that the ionization fraction $x$
is well approximated by the electron abundance $x_e$
shown in the third panel of Figure \ref{fig:profileprops}. From that figure
we see that the ionization fraction remains small, $x \la 0.02$, within
the cloud, even for the highest X-ray flux level. Thus $R$ in Equation
\ref{eq:tion} is nearly constant, $R \sim 2.6 \times 10^{-18}$\,s$^{-1}$.

Figure \ref{fig:comparetimes} shows that for all the cases with a soft UV
Lyman-Werner background, i.e. with or without an extra ionizing X-ray
component, the pregalactic cloud is
in photodissociation equilibrium over much of its interior
($10 \la r \la 100$~pc). Thus the \HH number
density is well approximated by its photodissociation equilibrium value
\begin{equation}
n_{H2}^{equil} \approx \frac{k_7 n_{HI} n_e}{k_{diss}}.
\label{eq:photodiss}
\end{equation}
The photodissociation timescale is much shorter than either the recombination
or ionization times in this region.  In addition, for the cases with the most
X-ray flux, the cloud is also close to ionization equilibrium. However, the
recombination rate is still slightly faster than the ionization rate. 
We can use the fact that the cloud is both in photodissociation 
equilibrium and in approximate ionization equilibrium
over these radii ($10 \la r \la 100$~pc) to understand the scaling
properties of $n_{H2}$ seen in the bottom panel of Figure 
\ref{fig:profileprops}.  Since $n_{HII} \approx n_e$ ionization equilibrium 
implies that the electron number density $n_e^{equil}$ scales as
\begin{equation}
n_e^{equil} \propto (\epsilon_x F_{LW} n_{HI})^{1/2}T^{0.32}
\label{eq:neioneq}
\end{equation}
Thus for interior regions where both ionization and photodissociation 
equilibrium hold we may substitute Equations \ref{eq:kdiss} and 
\ref{eq:neioneq} and into Equation \ref{eq:photodiss} (for fixed $F_{LW}$) 
to obtain 
\begin{equation}
n_{H2}^{equil} \propto \epsilon_x^{1/2}n_{HI}^{3/2}T^{1.2}
\label{eq:equilscale}
\end{equation}
Since at these radii Figure \ref{fig:profileprops} shows that 
the density and temperature profiles are similar, we expect the amount of 
\HH coolant to scale roughly as $n_{H2} \sim \epsilon_x^{1/2}$. 
Thus changing the X-ray flux by a factor
$100$ only changes the amount of \HH coolant by about a factor of $10$ 
(as seen in the bottom panel of Figure \ref{fig:profileprops}). 
Only in the core region do we see ionization dominate for the high X-ray flux
cases; however by this redshift (see Figure \ref{fig:profileprops}) the
core has already maximally cooled. X-ray enhanced production of molecular
hydrogen in this region will not promote additional cooling. Thus the 
positive feedback effect of X-rays, while present, is too slow to 
dramatically reverse the delay in cooling and collapse caused by the 
rapid photodissociation of \HH in these systems.
Note also that the dynamical collapse time
is faster than recombination or ionization times over most of cooling region
for $\epsilon_x=0.1$ and comparable to $t_{rec}$ and $t_{dyn}$ for
$\epsilon_x=1$, so density evolution can not be ignored.

\begin{figure*}
\includegraphics{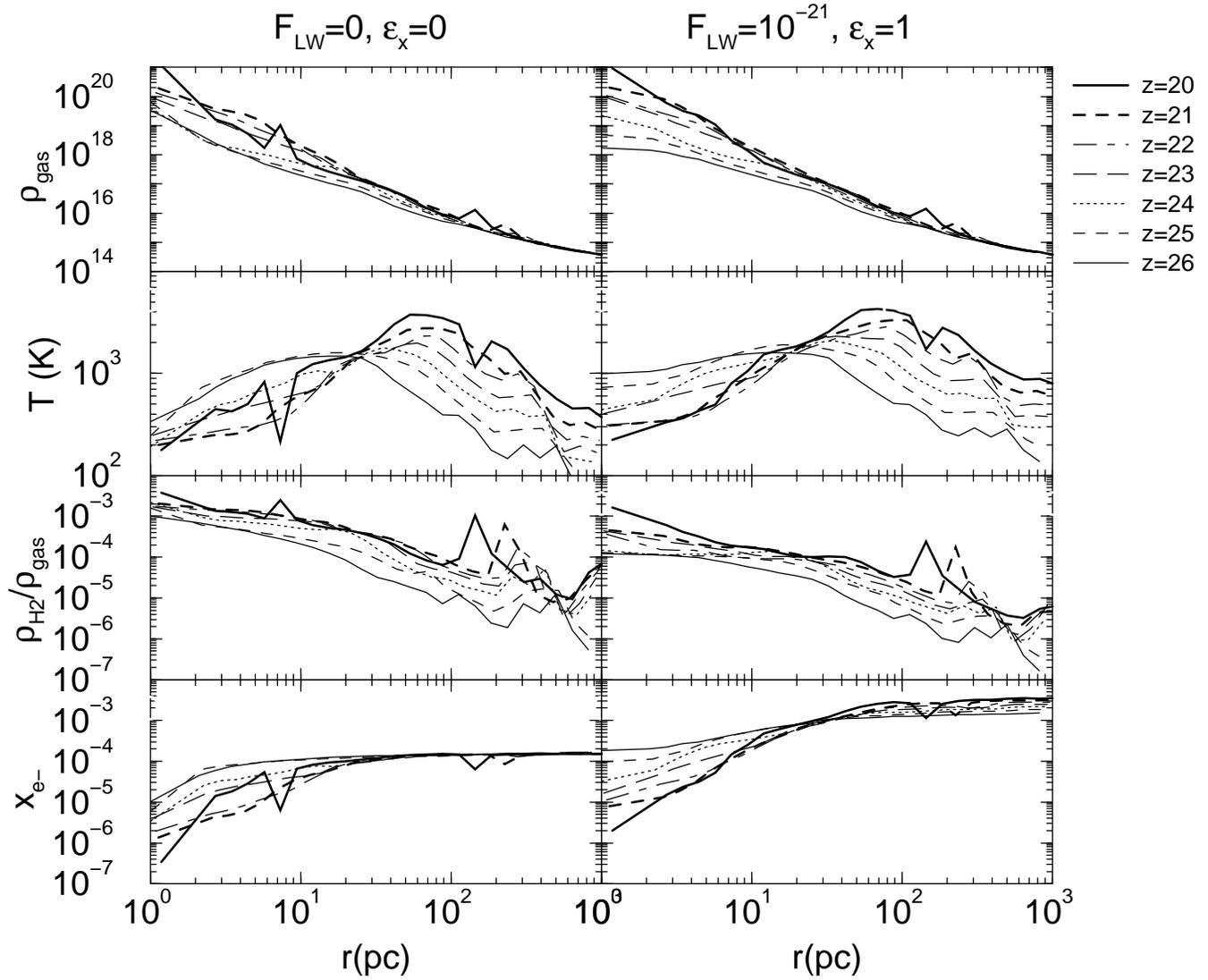}
\caption{Evolution of radial profiles of (from top to bottom)
the gas density $\rho_{gas}$, temperature $T$, \HH mass fraction
$\rho_{H2}/\rho_{gas}$ and
electron abundance $x_e$ from $z=26$ to $z=20$ for
the pregalactic cloud of Figure \protect\ref{fig:profileprops}.
Left panels show the case with no background radiation fields
present ($F_{LW}=0$, $\epsilon_x=0$);
while right panels show the case with maximal positive feedback
($F_{LW}=10^{-21}$~\fluxunit, $\epsilon_x=1$). Lines denote redshifts
$z=26$(thin solid), $25$(thin dashed), $24$(dotted), $23$(long dashed), 
$22$(dot dashed), $21$(thick dashed), and $20$(thick solid).}
\label{fig:evolveprops}
\end{figure*}

We can check that this is actually the case for our collapsing cloud by
tracking the evolution of its internal properties.
In Figure \ref{fig:evolveprops} we show the
evolution of the density $\rho_{gas}$, temperature $T$, \HH mass fraction
$\rho_{H2}/\rho_{gas}$, and electron abundance $x_e$ from redshifts
$26 \ge z \ge 20$ for the case with maximal positive feedback
$\epsilon_x=1$ (right panels) and the case with no background radiation
fields $F_{LW}=0$,$\epsilon_x=0$ (left panels).  For the $\epsilon_x=1$ case
at high redshift $z \sim 26$, the electron fraction is not significantly
enhanced in the core region. The photodissociation timescale is much shorter
than all other timescales in the problem so that the level of \HH coolant is
controlled by its photodissociation equilbrium value. At this redshift the
cloud temperature is still high ($\sim 1000$~\kel), the \HH mass fraction is
near its critical value (see MBAI) and the cloud has just begun to cool. The
density profile in the core is still roughly constant, as expected for a cloud
before collapse, and gradually steepens into a characteristic $r^{-2}$ form
by $z \sim 23$.  In contrast, the control case with no radiative feedback
evolves much more quickly. By $z=26$ the fraction of molecular hydrogen,
whose build up in this case is not regulated by photodissociation, is
$\ga 10^{-3}$, more than an order of magnitude greater than for the
$\epsilon_x=1$ case at the same redshift. The temperature in the core is
$\sim 450$~\kel at $z = 26.5$ and has cooled to its minimum value by
$z =26$. The cloud has collapsed as indicated by the steep core density
profile.  Densities in the core are more than an order of
magnitude greater than those for the case with radiative feedback. We also see
from Figure \ref{fig:evolveprops} that the electron abundance near the virial
radius is an order of magnitude greater for the case with X-rays than without.
Thus the dominant effect of the X-rays is to partially ionize the lower
density regions. We also see in both cases evidence for and the importance of
growth through merging of smaller substructures in the formation of the cloud.

In the outer low density region $r \ga r_{vir}$, we see enhanced \HH
formation caused by the increased electron fraction; however, the levels of
\HH remain below the critical threshold (see MBAI) for cooling to be
important.  We also caution the reader that the \HH fractions for the
low density regions shown in Figure \ref{fig:profileprops} and the
right panel of Figure \ref{fig:evolveprops}
should be considered conservative  upper limits on the amount of coolant
present because H$^-$ photodetachment, which we ignore, is no longer
negligible once $n_{HI} \la 0.045$
($\rho_{gas} \la 6 \times 10^{15}\, \MsMpc$).
We find that the dominant effect of the X-rays at these large radii is to
heat and partially ionize the intergalactic medium, in qualitative agreement
with recent work on the IGM by Venkatesan \etal (2001).

\section{Summary}
\label{sec:conclude}

In this paper we used high resolution numerical simulations to investigate
the effect of radiative feedback on the formation of $10^5$ --$10^7 \Ms$
pregalactic clouds when the radiation spectrum extends to energies
above the Lyman limit ($ \ga 1$~keV). Such an ionizing X-ray component is
expected if the initial mass function of the first luminous sources contains
an early generation of miniquasars or very massive stars. The range of
pregalactic objects we consider is important because they are large enough to
form molecular hydrogen, but too small to cool by hydrogen line cooling. Thus
any process that affects the amount of \HH coolant within the cloud affects
its ability to cool, lose pressure support and collapse to high density.
The soft, UV flux in the $11$--$13$~eV Lyman-Werner band produced by the
first stars can destroy the fragile \HH in these objects delaying
subsequent collapse and star formation until later redshifts when the
objects have evolved to larger masses.  We test whether the presence of
ionizing X-rays can mitigate or even reverse this effect by
increasing the electron fraction in the gas and
thus enhancing the formation of molecular hydrogen coolant. Since the
relative amplitude of the X-ray to soft UV components in the background
spectrum of the first luminous sources is unknown, we study four cases with
 relative X-ray normalizations ranging from zero to ten for mean soft UV flux
at $12.86$~eV of $10^{-21}$~\fluxunit. We compare these results to
the case with no background radiation fields.  We draw our initial
conditions from a \LCDM cosmological model. The simulations evolve the
nonequilibrium rate equations for $9$ species of hydrogen and helium including
the effects of secondary electrons.
 A summary of our main findings are as follows:
\begin{itemize}
\item Ionizing X-rays do have a positive effect on subsequent structure
formation, but the effect is very mild. Even in the presence of X-rays
photodissociation is rapid delaying the collapse of the cloud until later
redshifts when larger objects have collapsed.
\item The mass thresholds for gas to cool and for gas to cool and
become dense decrease only weakly with increasing
X-ray flux up to relative X-ray normalization $\epsilon =1$ when compared
to the case with only a soft UV radiation field, but remain
 $ \ga$ a factor three more massive than the mass collapse threshold found
when no radiation fields were present. Equivalently, the redshift for collapse
decreases weakly with increasing X-ray flux up to $\epsilon_x = 1$ from that
for the case with only a soft UV radiation field, but collapse
occurs significantly later than in the case with no background radiation
field.
\item The fraction of gas that can cool or cool and become dense (and thus
become available for star formation) within a cloud increases with increasing
X-ray flux.  We fit these fractions with a simple fitting formula (Equation
\ref{eq:regression}) that increases logarithmically with cloud mass and find
that the slope of this fitting formula increases by as much as a factor
$\sim 2$ with increasing X-ray flux for X-ray normalizations
$\epsilon_x \le 1$.
\item The weak positive effect of the ionizing X-rays appears maximal for
relative normalization $\epsilon_x=1$. For significantly higher
X-ray fluxes the positive trends described in the previous two items is
reversed. Heating becomes important both within the cloud and in the
surrounding intergalactic medium thus weakening the characteristic
accretion shock near the virial radius. The mean temperature of the cloud is
raised well above its virial temperature causing a significant fraction of
the gas to be evaporated into the surrounding intergalactic medium.
\end{itemize}

We conclude that although an early X-ray background from quasars or
mini-quasars does enhance cooling in pregalactic objects, the effect
is weaker than found in previous studies that did not follow the
evolution of the collapsing cloud. The net impact on subsequent
structure formation is still negative due to photodissociation of the
\HH coolant by the soft UV radiation spectrum of the first stellar
sources, and the pattern of subsequent structure formation is only
weakly changed by including an ionizing X-ray component.
We should point out that that we have not included star formation and
its subsequent effect on the forming halos in these simulations
(except insofar as we have modelled the radiative background), so
obviously more work on this subject is required.  In this context, it
is interesting to note that Ricotti \etal (2002a, 2002b) have simulated
star formation and radiative transfer in somewhat more massive halos
(albeit with a mass resolution 100 times lower than used here) and
find a self-regulated feedback loop that includes positive feedback.


Together with the findings of MBAI our results suggest that radiative
feedback from cosmological radiation backrounds have subtle effects on
the formation of luminous objects within the micro--galaxies. The
negative feedback of a soft UV background may change the minimum mass
of a dark halo within which gas may cool by a factor of a
few. However, even in the most extreme cases considered the first
objects to form rely on molecular hydrogen as coolant. Hence, our
results do {\sl not} justify the neglect of halos cooling by molecular
hydrogen in all current studies of galaxy formation.

We found that heating from an early X-ray background only slightly
modifies the temperature and density profiles of halos at the time
when a cool core is first formed in their centers. One may speculate
that such temperature variations may lead to varying accretion rates
onto the proto--star which will form within them (Abel, Bryan \&
Norman 2002). If so early radiation backgrounds may influence the
spectrum of initial masses of Population III stars. To answer such
detailed questions will rely on carrying out yet higher resolution
simulations than the ones presented here. 



This work is supported in part by National Science Foundation grant
ACI-9619019.  The computations used the SGI Origin2000 at the National
Center for Supercomputing Applications. M.E.M. gratefully
acknowledges the hospitality and support of the MIT Center for Space
Research where most of this work was done.



\newpage

\clearpage

\clearpage
\begin{table}
\centerline{\begin{tabular}{|c|c|c|c|c|c|c|} \hline
$\epsilon_x$ & $M_{vir}$ & $M_{gas}$ & $M_{gas}/M_{vir}$ & $T_{mean}$ &
 $T_{vir}$ \\
   &$10^6\,\Ms$  & $10^6\,\Ms$ &  & K  & K   \\
\hline
\hline
$0$ ls & $5.0$ & $0.50$ & $0.10 $ & $3200$ & $5300$   \\ \hline
$0$ gp & $5.0$ & $0.49$ & $0.10 $ & $3000$ & $5300$   \\ \hline
$0.1$  & $5.1$ & $0.48$ & $0.09 $ & $2500$ & $5400$   \\ \hline
$1$    & $5.1$ & $0.47$ & $0.09 $ & $2200$ & $5400$   \\ \hline
$10$   & $4.9$ & $0.37$ & $0.07 $ & $3800$ & $5200$   \\ \hline
none   & $5.0$ & $0.43$ & $0.09 $ & $1700$ & $5300$   \\ \hline
\end{tabular}}
\caption{Mean properties of the most massive peak at redshift $z=20$
exposed to a mean Lyman-Werner flux $F_{LW}=10^{-21}$~\fluxunit and
ionizing X-ray background whose spectrum is given by Equation
\protect\ref{eq:spectrum} where $\epsilon_x$ denotes the relative X-ray
flux normalization. The row labeled ``none'' denotes the $F_{LW}=0$,
$\epsilon_x=0$ case.}
\label{tab:meanprops}
\end{table}

\begin{table}
\centerline{\begin{tabular}{|c|c|c|c|c|c|c|} \hline
$\epsilon_x$ & $B_c$ & $M_c$ & $K_c$ & $B_{cd}$ & $M_{cd}$ & $K_{cd}$ \\
             &       & $10^5 \Ms$ &  &  &$10^5 \Ms$  &  \\
\hline
\hline
$0$ ls & $0.077$ & $8.5$ & $0.87$ & $0.033$ & $10$  & $0.83$  \\ \hline
$0$ gp & $0.113$ & $8.6$ & $0.86$ & $0.047$ & $9.5$ & $0.84$  \\ \hline
$0.1$  & $0.171$ & $7.9$ & $0.87$ & $0.084$ & $8.8$ & $0.79$  \\ \hline
$1$    & $0.205$ & $7.0$ & $0.90$ & $0.111$ & $8.0$ & $0.79$  \\ \hline
$10$   & $0.166$ & $6.9$ & $0.85$ & $0.105$ & $8.3$ & $0.80$  \\ \hline
none   & $0.153$ & $2.7$ & $0.94$ & $0.080$ & $2.7$ & $0.82$  \\ \hline
\end{tabular}}
\caption{ Coefficients determined by a mean regression analysis assuming
the functional form $f_i=B_i \ln (M/M_i)$ ($M > M_i$) where $f_i$ are the
fractions of cold gas ($i=c$) and cold, dense gas ($i=cd$) formed in
pregalactic clouds when both soft UV Lyman-Werner and ionizing
X-ray backgrounds are present. The mean Lyman-Werner flux is
fixed at $F_{LW}=10^{-21}$~\fluxunit. Relative X-ray normalizations
are given by $\epsilon_x$, $K_i$ are the respective correlation
coefficients for the fits, ls (gp) denotes the use of
Lepp \& Shull (Galli \& Palla) cooling functions, and ``none'' labels
the $F_{LW}=0$, $\epsilon_x=0$ case.}
\label{tab:coldfracfits}
\end{table}

\end{document}